\documentclass[]{raa}            
\usepackage{graphicx,times}
\usepackage{natbib}

\providecommand{\bm}[1]{\mbox{\boldmath$#1$\unboldmath}}

\begin{document}

\newcommand{\beq}[1]{\begin{equation}\label{#1}}
 \newcommand{\eeq}{\end{equation}}
 \newcommand{\bea}{\begin{eqnarray}}
 \newcommand{\eea}{\end{eqnarray}}
 \def\disp{\displaystyle}

   \title{Testing the Distance-Duality Relation with a Combination
   of Cosmological Distance  Observations
}

 \volnopage{ {\bf 2011} Vol.\ {\bf 11} No. {\bf XX}, 000--000}
   \setcounter{page}{1}

   \author{Shuo Cao \inst{1}
   \and Nan Liang \inst{1,2}
        }


   \institute{Department of Astronomy, Beijing Normal University, Beijing 100875, China; {\it liangn@bnu.edu.cn}\\
     \and
   Center for High Energy Physics, Peking University, Beijing 100871, China\\
\vs \no
   {\small Received [year] [month] [day]; accepted [year] [month] [day] }
}

\abstract{ In this paper, we propose an accurate test of the
distance-duality (DD) relation, $\eta=D_{L}(z)(1+z)^{-2}/D_{A}(z)=1$
(where $D_{L}$ and $D_{A}$ are the luminosity distances and angular
diameter distances, respectively), with a combination of
cosmological observational data of Type Ia Supernave (SNe Ia) from
Union2 set and the galaxy cluster sample under an assumption of
spherical model. In order to avoid bias brought by redshift
incoincidence between  observational data and to consider redshift
error bars of both clusters and SNe Ia in analysis, we carefully
choose the SNe Ia points which have the minimum acceptable redshift
difference of the galaxy cluster sample ($|\Delta
z|_{\rm min}
=\sigma_{z, \rm SN}+\sigma_{z, \rm cluster}$). By assuming $\eta$ a
constant and functions of the redshift parameterized by six
different expressions, we find that there exists no conceivable
evidence for variations in the DD relation concerning with
observational data, since it is well satisfied within $1\sigma$
confidence level for most cases. Further considering different
values of $\Delta z$ in constraining, we also find that the choosing
of $\Delta z$ may play an important role in this model-independent
test of the
distance-duality relation for the spherical sample of galaxy clusters. 
 }

\authorrunning{Shuo Cao and Nan Liang}
\titlerunning{Testing the Distance-Duality Relation}
   \maketitle


\section{Introduction}
\label{sec:introduction}

Distance-duality relation, also known as the Etherington's
reciprocity relation \citep{eth33}, is of fundamental importance in
cosmology, which relates the luminosity distance (LD, $D_{L}$) with
the angular diameter distance (ADD, $D_{A}$) by means of the
following expression,
\begin{equation}
  \eta =\frac{D_{\scriptstyle L}}{D_{\scriptstyle A}}{(1+z)}^{-2}=1.
  \label{rec1}
\end{equation}
We notice that the DD relation is completely general, valid for all
cosmological models based on Riemannian geometry, being dependent
neither on Einstein field equations nor the nature of matter-energy
content. It only requires that source and observer are connected by
null geodesics in a Riemannian spacetime and that the number of
photons are conserved. This equation plays an essential role in
modern cosmology  \citep{Csaki02}, ranging from gravitational
lensing studies \citep{Schneider92,Fu08} to analyzes from galaxy
clusters observations \citep{Lima03,Cunha07}, as well as to the
plethora of cosmic consequences involving primary and secondary
temperature anisotropies of the cosmic microwave background (CMB)
observations \citep{Komatsu10}.

Up to now, diverse astrophysical mechanisms such as gravitational
lensing and dust extinction have been proved to be capable to cause
obvious deviation from the distance duality and testing this
equality with high accuracy can also provide a powerful probe of
exotic physics \citep{Bassett04a,Bassett04b,Corasaniti06}.
Therefore, it is rewarding to explore the distance-duality relation
to test the validity of photon conservation etc. On the side of the
observational data, if one is able to find cosmological sources
whose intrinsic luminosities are known (standard candles) as well as
their intrinsic sizes (standard rulers), one can determine both
$D_{\scriptstyle L}$ and $D_{\scriptstyle A}$, and after measuring
the common redshifts, to test directly the above Etherington's
result. The possibility of using the Sunyaev-Zeldovich effect (SZE)
together with X-ray emission of galaxy clusters to measure angular
distances was suggested soon after the SZE was found \citep{Silk78}.
Using jointly,  an independent method to determine distance scales
and thus to measure the value of the Hubble constant has been
provided \citep{Silk78,Birkinshaw91,Inagashi95,Nozawa06}. By using
an isothermal spherical model for which the hydrostatic equilibrium
model and spherical symmetry assumed, \cite{Reese02} selected 18
galaxy cluster sample and \cite{Boname06} obtained 38
ADD galaxy clusters sample. 
 \cite{Filippis} have
corrected the samples  by using an isothermal elliptical model to
get 25 ADDs of galaxy clusters. \cite{Uzan04} considered 18 ADDs
sample \citep{Reese02} to test the DD relation by assuming the
$\Lambda$CDM model via the technique,
$D_A^{\mathrm{cluster}}(z)=D_A^{\Lambda \rm CDM}(z)\eta^2(z)$. They
showed that no violation of the DD relation is only marginally
consistent. Using the DD relation for astrophysical research can be
found in many works, e. g.,
\cite{Bassett04b,More09,Avg10,HLR11a,Cao11}.

In order to test the DD relation in a model-independent way, one
should use measurements of $D_L$ from cosmological observations
directly. The first direct evidence for cosmic acceleration came
from Type Ia Supernave (SNe Ia) \citep{Riess98,Perlmutter99}, which
have provided the strongest constraints on the cosmological
parameters \citep{Riess04,Riess07,Astier06,Wood07,Kowalski08}.
\citet{bem06} divided the weighted average of galaxy clusters
\citep{Boname06} and SNe Ia \citep{Riess04} in redshift bins and
concluded that the validity of $\eta = 1$ is consistent at $68.3\%$
(1$\sigma$) CL. Recently, \cite{HLR10} tested the DD relation with
ADD samples and the Constitution set of SNe Ia data
\citep{Hicken09}. In order to avoid the corresponding bias of
redshift differences, a selection criteria ($\Delta z=|z_{\rm
SN}-z_{\rm cluter}|\le0.005$) for a given pair of data set are used.
By using two parameterizations of $\eta$ parameter, they found that
the DD relation are marginally compatible within $2\sigma$ CL with
the elliptical model sample \citep{Filippis}, and a strong violation
($>3\sigma$) of the DD relation with the spherical model sample
\citep{Boname06}. However, \citet{Li11} found that, by removing more
data points of galaxy clusters samples according  to the selection
criteria, the DD relation can be accommodated at $1\sigma$ CL for
the elliptical model, and at $3\sigma$ CL for the spherical  model.
\cite{Nair11} discussed the validity of the DD relation  with
observational data and ruled out some of the parameterizations
significantly.

It is obvious that the difference of redshifts between galaxy
clusters and SNe Ia may cause obvious deviation in testing the DD
relation. In principle, the only strict criterion to form a given
pair is that galaxy clusters and SNe Ia should be at the same
redshift. \cite{Liang11} have found that the DD relation is
satisfied at $1\sigma$ CL with the corrected $D_{L}$ located at the
same redshift of the corresponding 38 spherical galaxy cluster
sample, which are obtained by interpolating from the nearby SNe Ia
of the Union2 set. It should be noted that the redshifts of
observations are not determined with infinite accuracy, and there is
no point to decrease $\Delta z$ below the total $1\sigma$ error of
observational redshifts $\sigma_{z, \rm tot}=\sigma_{z,\rm
SN}+\sigma_{z,\rm cluster}$. Therefore, the finite errors of both
clusters and SNe Ia should be taken into account in the analysis. In
this paper,  we consider redshift error bars of both clusters and
SNe Ia in analysis of avoiding bias from redshift differences
between observational data to test the DD relation. In practice,
$\sigma_{z,\rm tot}$ is not smaller than 0.002, therefore it is not
appropriate to use a smaller window constraint. For the total 38
data pairs with the spherical sample of galaxy clusters and the
Union2 set, we find that differences of redshifts between total 38
data pairs are very small ($\Delta z\le0.005$), and there are 33
pairs for the minimum selection criteria $|\Delta z|_{\rm
min}=\sigma_{z,\rm tot}$. Thus we choose the SNe Ia points which
have the minimum acceptable redshift difference of the galaxy
cluster sample $\Delta z\le0.002$. This criteria serves a much more
stringent one compared with $\Delta z\le0.005$ \citep{HLR10,Li11},
therefore the accuracy and reliability of our test should be
improved. We also find that the choosing of $\Delta z$ may play an
important role in this model-independent test.

This paper is organized as follows. In Section~\ref{sec:model}, we
introduce seven parametrizations for the DD relations applied in
this work. In Section~\ref{sec:Sample}, we present a combined data
given by the latest released Union2 SNe Ia data as well as the 38
galaxy cluster samples under an assumption of spherical model. In
Section~\ref{sec:analysis}, we briefly describe the analysis method
and show results for constraining parameters of the DD relation.
Finally, we summarize conclusions in Section~\ref{sec:Conclusions}.

\section{DD relation parametrizations } \label{sec:model}

Regarding the parametrization of the DD relation, a model
independent test has been extensively discussed in the above quoted
papers \citep{bem06,HLR10,HLR11a,HLR11b,Li11,Nair11,Liang11,Meng11}.
\citet{bem06} considered $\eta$ a constant with no relation to the
redshift \\

\hspace{1.0cm} I. $\eta = \eta_{0} $, \\
where $\eta_0$ is a constant to be constrained by observational
data. In general,  $\eta$ can be treated as parameterized functions
of the redshift, $\eta(z)$, which are clearly inspired on similar
expressions for $w(z)$, the equation of state in time-varying dark
energy models (see, for instance,
\citet{CP01,Linder03,Cunha07,Silva07}). Recently,
\citet{HLR10,HLR11a}
used two one-parameter  expressions, namely,\\

\hspace{1.0cm} II. $\eta (z) = 1 + \eta_{a} z$,

\hspace{1.0cm} III. $\eta (z) = 1 + \eta_{a}z/(1+z)$ \\
In this work, we also use another general parametric representations
for a possible redshift dependence of the distance
duality expression including three two-parameter parameterizations \citep{Li11,Nair11,Liang11,Meng11} \\

\hspace{1.0cm} IV. $\eta (z) = 1 + \eta_{a}\ln(1+z)$,

\hspace{1.0cm} V. $\eta (z) = \eta_0 + \eta_{a}z$,

\hspace{1.0cm} VI. $\eta (z) = \eta_0 + \eta_{a}z/(1+z)$,

\hspace{1.0cm} VII. $\eta (z) = \eta_0 + \eta_{a}\ln (1+z)$.

\section{Galaxy Clusters and Supernovae Ia Samples }\label{sec:Sample}

In this work, we consider the sample of ADD from galaxy clusters
obtained by combining their SZE and X-ray surface brightness
observations  sample \citep{Boname06}. Under an assumption of
spherical model, the cluster plasma and dark matter distributions
were analyzed assuming hydrostatic equilibrium model and spherical
symmetry, thereby accounting for radial variations in density,
temperature and abundance. Recently, the Supernova Cosmology Project
(SCP) collaboration have released their Union2 compilation which
consists of 557 SNe Ia \citep{Amanullah}, which is the largest
published and spectroscopically confirmed SNe Ia sample to date.

For a given $D_A^{{\rm cluster}}$ data point, theoretically, we
should select an associated SNe Ia data point $D_L^{{\rm SN}}$ at
the same redshift to obtain an $\eta_{\rm obs}$. In order to avoid
any bias of redshift differences between SNe Ia and galaxy clusters
and to consider redshift error bars of both clusters and SNe Ia in
analysis, we should determine the value of $\sigma_{z,\rm
tot}=\sigma_{z,\rm SN}+\sigma_{z,\rm cluster}$ for the combination
of observational data pairs. For the observations of SNe Ia, the
peculiar velocity uncertainty is set at $400 \rm km s^{-1}$
\citep{Wood07} (or $300 \rm km s^{-1}$, \cite{Kowalski08}) and the
redshift uncertainty is $\sigma_{z,\rm SN}=0.001$ \citep{Hicken09}.
For the observations of galaxy clusters, the rms one-dimensional
cluster peculiar velocity uncertainty is set at $256^{+106}_{-75}\rm
km s^{-1}$, which corresponds to the three-dimensional rms velocity
$459^{+184}_{-130}\rm km s^{-1}$ \citep{Watkins97} (or $341\pm 93\rm
km s^{-1}$ for the rms one-dimensional cluster peculiar velocity,
which corresponds to the three-dimensional rms velocity $591\pm
161\rm km s^{-1}$, \cite{Dale99}) and the redshift uncertainty is
$\sigma_{z,\rm cluster}=0.001$. Therefore, $\Delta z=\sigma_{z,\rm
tot}=0.002$ is considered in our work. Obviously, this strict choice
with $\Delta z =0.002$ may helpfully ease the systematic errors
brought by redshift inconsistence between SNe Ia and galaxy
clusters. Therefore, we obtain a sub-sample of SNe Ia from the
Union2 data set whose redshifts coincide with the ones appearing in
the galaxy cluster sample under this criterion. We then bin the SNe
data in the redshift bins of the corresponding spherical galaxy
cluster sample to obtain 33 data pairs in our test. Assuming that
$\mu_i$ represents the $ith$ appropriate SNe Ia distance modulus
data (within the $|\Delta z|<0.002$ redshift range) with
$\sigma_{\mu_i}$ denoting its reported observational uncertainty, in
light of standard data reduction framework by
\citet[Chap.~4]{Bev03}, we obtain
\begin{equation}
\begin{array}{l}
\bar{\mu}=\frac{\sum\left(\mu_i/\sigma^2_{\mu_i}\right)}{\sum1/\sigma^2_{\mu_i}},\\
\sigma^2_{\bar{\mu}}=\frac{1}{\sum1/\sigma^2_{\mu_i}},
\end{array}
\end{equation} where $\bar{\mu}$ stands for the weighted mean
distance modulus at the corresponding galaxy cluster redshift and
$\sigma_{\bar{\mu}}$ serves as its uncertainty.

It must be emphasized that, if a redshift-dependent expression for
the DD relation is considered, the SZE+X-ray surface brightness
observation technique gives
$D_A^{\mathrm{cluster}}(z)=D_A(z)\eta^2(z)$ \citep{Sunyaev,
Cavaliere}. So, we must replace $D_A(z)$ with
$D_A^{\mathrm{cluster}}(z)\eta^{-2}$ when we try to test the
reciprocity relation  consistently with the SZE+X-ray observations
from galaxy clusters. Thus, the observed $\eta_{\rm obs}(z)$ is
determined by the following expression:
\begin{eqnarray}\eta_{\rm obs}(z)=D_A^{\mathrm{cluster}}(z)(1+z)^2/D_L(z).\end{eqnarray}

It should be noted that the data points of the compiled Union2 SNe
Ia are given in terms of the distance modulus, which could reduce to
\begin{equation}
  D_L(z)=10^{\mu(z)/5-5}.
  \label{dl}
\end{equation}
Accordingly, the uncertainty of the luminosity distance could be
expressed in term of the distance modulus uncertainty
$
\sigma_{D_L(z)}=\ln10/5\times 10^{(\mu/5-5)}\sigma_{\mu(z)}.
\label{errordl}
$

\section{Analysis and Results}\label{sec:analysis}

In this section, we estimate the $\eta_{0}$ and $\eta_{a}$
parameters in seven parametrizations listed in Sec. \ref{sec:model}.
To estimate the model parameters of a given parameterized form, we
use the minimum $\chi^2$ estimator following standard route
\begin{equation}
\chi^2(z;\mathbf{p})=\sum_z\frac{[\eta_{\rm
th}(z;\mathbf{p})-\eta_{\rm obs}(z)]^2}{\sigma^2_{\eta_{\rm obs}}},
\end{equation}
where $\eta_{\rm th}$  represents the theoretical value of $\eta$
parameter with the parameter set $\mathbf{p}$, and $\eta_{\rm obs}$
associated with the observational technique with the error of
$\sigma_{\eta_{\rm obs}}$, which comes from the statical
contributions and systematic uncertainties of the galaxy clusters
and SNe Ia, as well as the redshifts
\begin{equation}
\sigma_{\eta_{\rm
obs}}=\sigma_{D_A}^{\mathrm{cluster}}(1+z)^2/D_L-D_A^{\mathrm{cluster}}\sigma_{D_L}(1+z)^2/D^2_L+2D_A^{\mathrm{cluster}}\sigma_{z}(1+z)/D_L.
\end{equation}
For the one-parameter models, one should expect the likelihood of
$\eta_0$ or $\eta_a$ peaks at $\eta_0=1$ or $\eta_a=0$ ($\Delta
\chi^2 $ minimizes at $\eta_0=1$ or $\eta_a=0$), in order to satisfy
the DD relation. As to the two-parameter models, one should expect
$\eta_0=1$ and $\eta_a=0$ to be the best-fit parameters in the
confidence contours, if it consists with photon conservation and no
visible violation of the DD relation.

\begin{figure}
\begin{center}
\includegraphics[width=0.45\hsize]{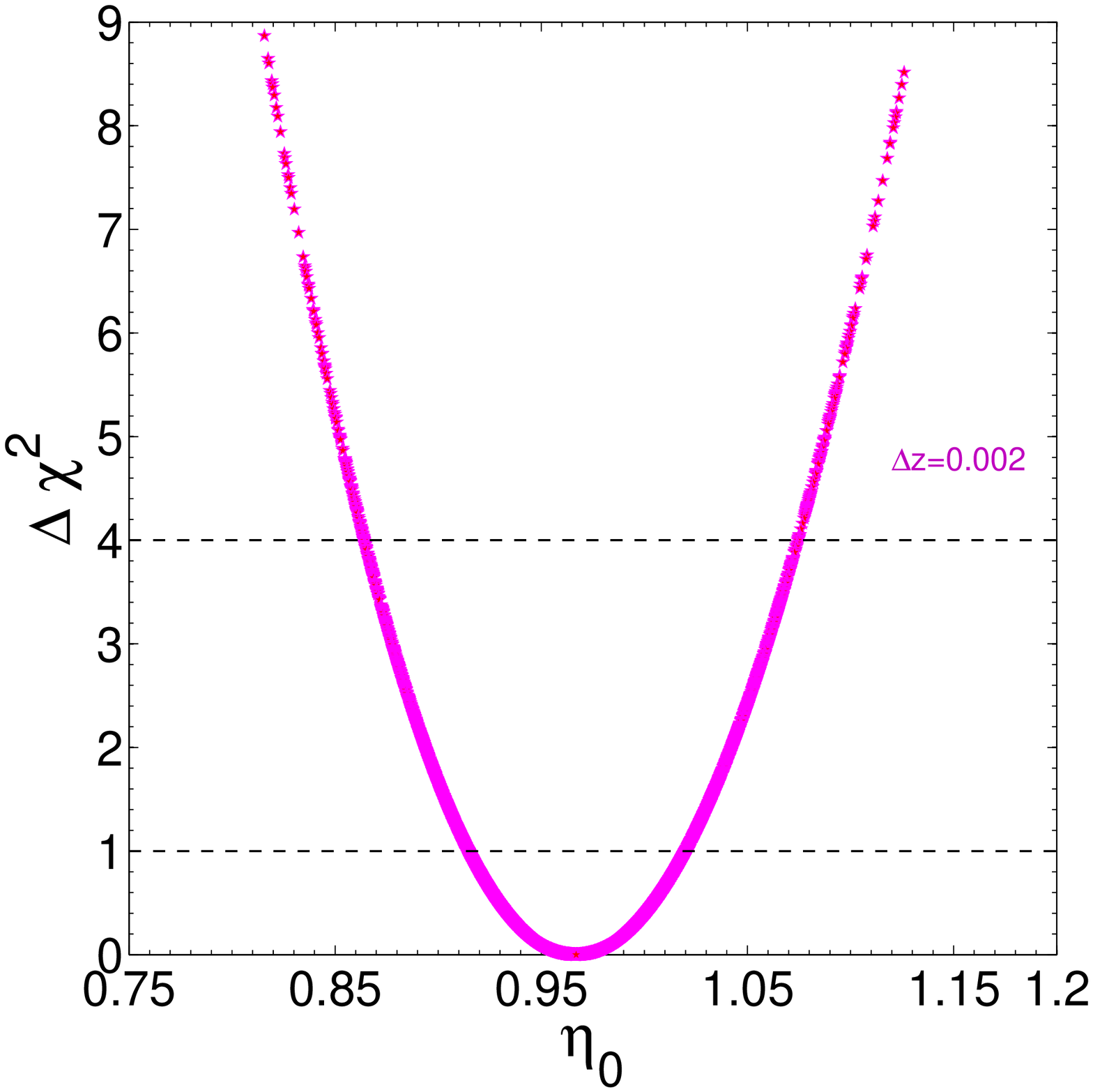}
\includegraphics[width=0.45\hsize]{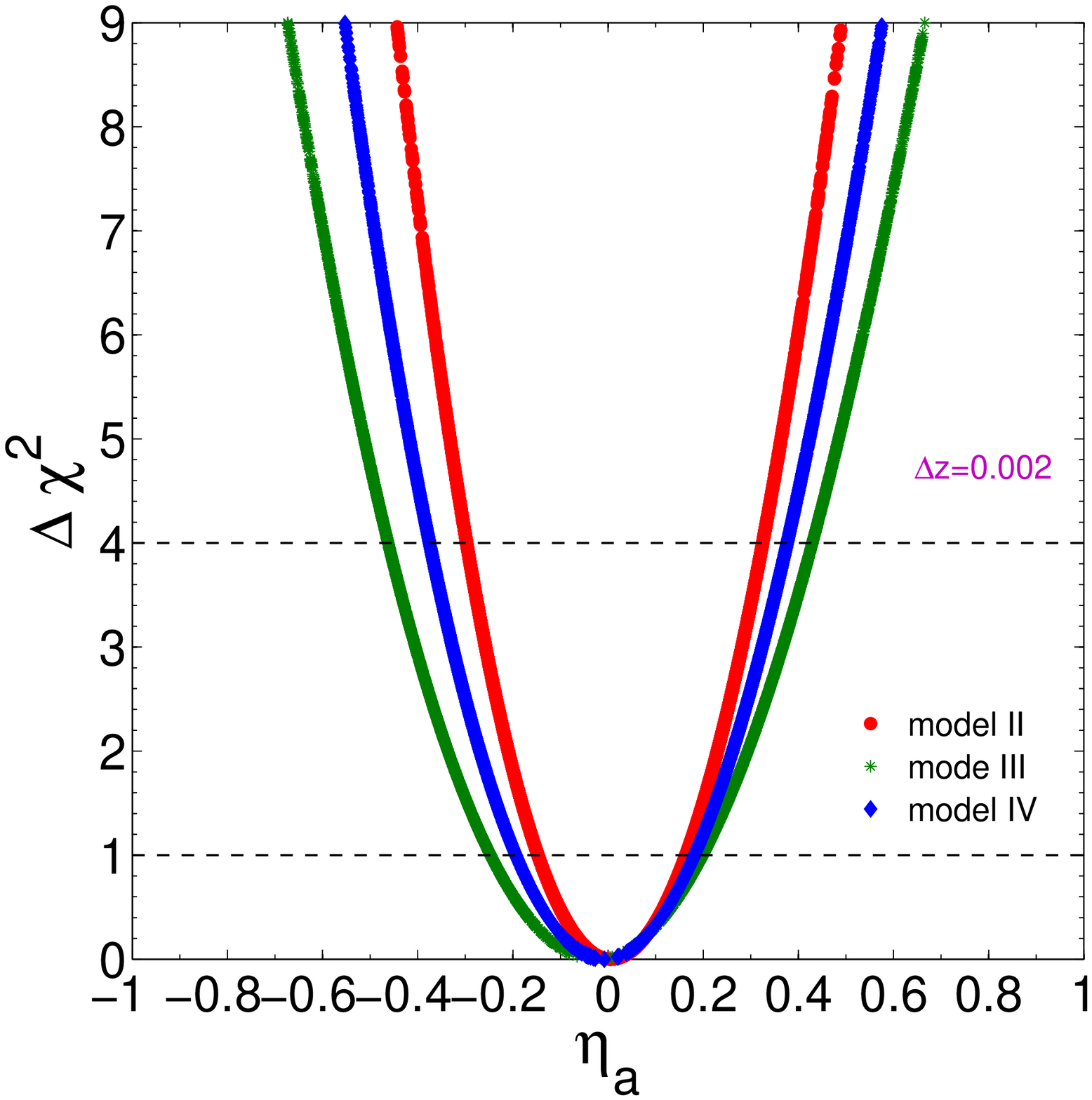}
\end {center}
 \caption{The likelihood distribution function with the 33
ADDs of galaxy clusters and the luminosity distances of the Union2
set for one-parameter forms in the $\eta_0-\Delta\chi^2$ planes (for
model I) and $\eta_a-\Delta\chi^2$ planes (for model II-IV) with
$\Delta z=0.002$.} \label{fig1}
\end{figure}

In Fig. \ref{fig1} (\textit{Left}), we plot the likelihood
distribution function in the $\eta_0-\Delta\chi^2$ plane and obtain
$\eta_{0}=0.97^{+ 0.05}_{-0.06}$ at 1$\sigma$, which is in good
qualitative accord with previous analyses ($\eta_{0} = 1.01^{+
0.07}_{-0.07}$) \citep{bem06}. In Fig.~\ref{fig1} (\textit{Right}),
we show the likelihood distribution function from three
one-parameter forms of the redshift: II. $\eta(z)=1+\eta_a z$; III.
$\eta(z)=1+\eta_a{z}/({1+z})$; and IV.  $\eta
(z)=1+\eta_{a}\ln(1+z)$. The best-fit values at $1\sigma$ CL are
$\eta_a=-0.01^{+ 0.15}_{-0.16}$ for model I., $\eta_a=-0.01^{+
0.21}_{- 0.24}$ for model II., and $\eta_a=-0.01^{+ 0.22}_{- 0.19}$
for model III., respectively, which are different from those
obtained in \citet{HLR10}, where the DD relation is ruled out at
$3\sigma$ CL, and those obtained in \citet{Li11}, where the DD
relation is accommodated at $3\sigma$ CL for the spherical model.
Fitting results from one-parameter forms with the ADDs of galaxy
clusters and the luminosity distances of the Union2 set with $\Delta
z=\sigma_{z,\rm tot}=0.002$ are summarized in Table \ref{tab1}.

\begin{table}[!h]\centering
\begin{tabular}{|l|c|c|}
\hline
Parameterization ($\Delta z=0.002$) ~&~~$\eta_0$ ~~&~~$\eta_a$ ~~\\
\hline
I. $\eta_0$  &~~$0.97^{+ 0.05}_{-0.06}$  ~~&~~ 0 ~~\\
\hline
II. $1+\eta_az$ &~~ 1 ~~&~~ $-0.01^{+ 0.15}_{-0.16}$ ~~\\
\hline
III. $1+\eta_a\frac{z}{1+z}$  &~~ 1~~&~~$-0.01^{+0.21}_{-0.24}$~~\\
\hline
IV. $1+\eta_a\ln(1+z)$  &~~ 1~~&~~$-0.01^{+ 0.22}_{- 0.19}$~~\\
\hline
V. $\eta_0+\eta_az$  &~~ $0.84^{+ 0.17}_{-0.17}$~~&~~$0.43^{+0.49}_{-0.49}$~~\\
\hline
VI. $\eta_0+\eta_a\frac{z}{1+z}$  &~~ $0.74^{+ 0.23}_{-0.22}$~~&~~$1.02^{+0.84}_{-0.85}$~~\\
\hline
VII. $\eta_0+\eta_a\ln(1+z)$  &~~ $0.82^{+0.20}_{-0.19}$~~&~~$0.57^{+0.68}_{-0.67}$~~\\
\hline
\end{tabular}
\tabcolsep 0pt \caption{\label{tab1} Summary of the results for
seven parameterizations at $1\sigma$ confidence level with $\Delta
z=0.002$.}  \vspace*{5pt}
\end{table}

\begin{figure}
\begin{center}
\includegraphics[width=0.33\hsize]{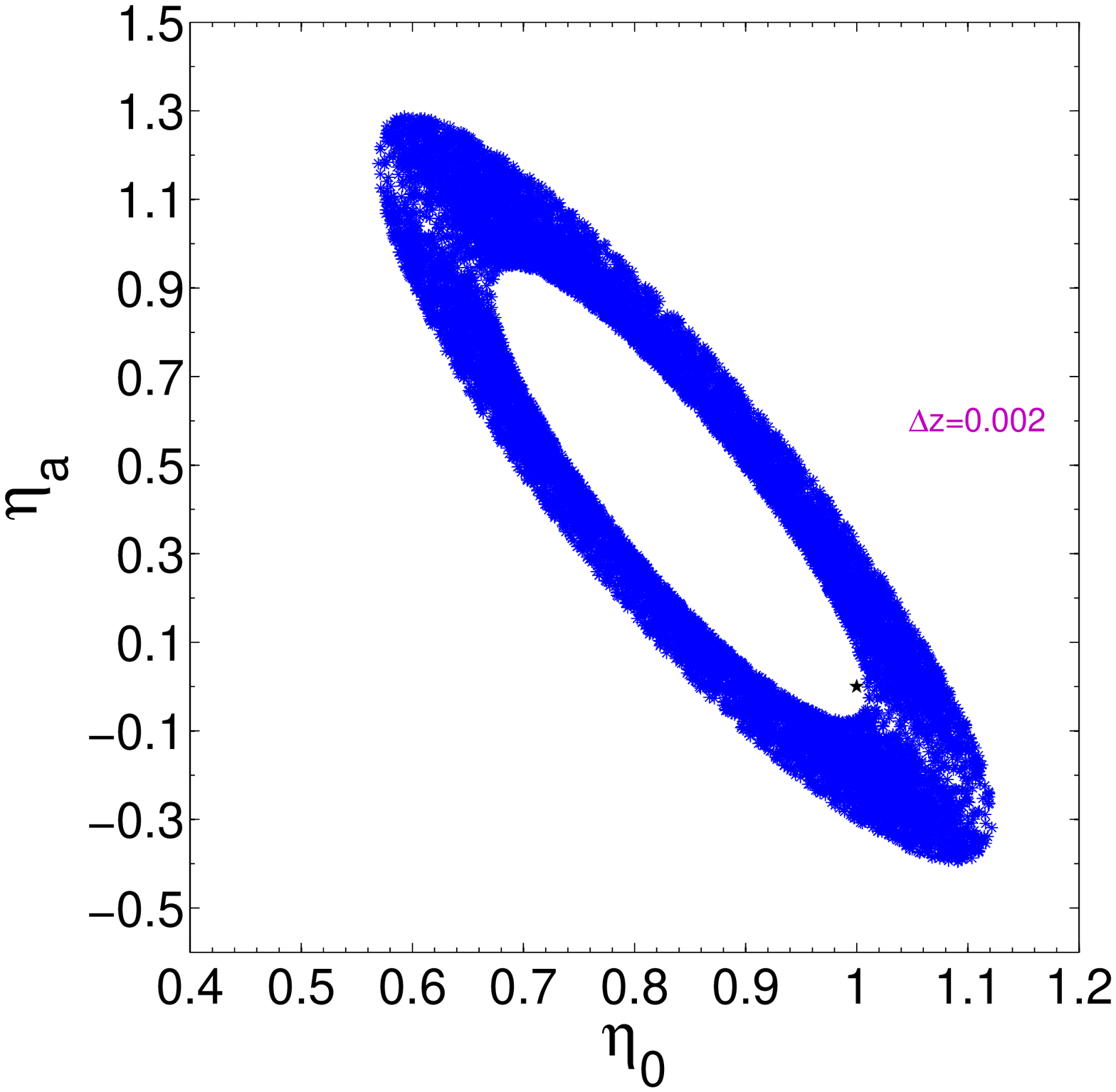}
\includegraphics[width=0.33\hsize]{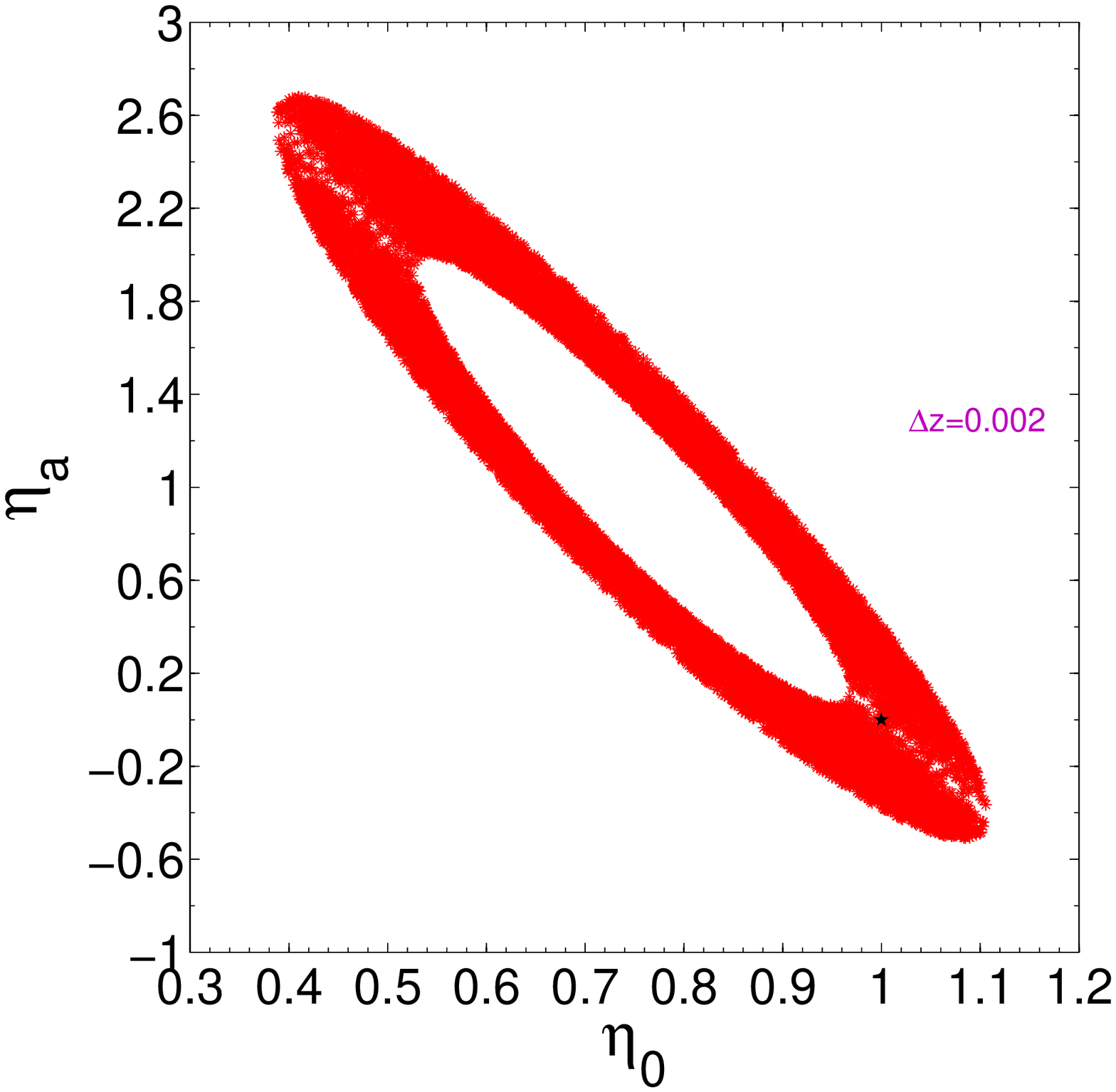}
\includegraphics[width=0.33\hsize]{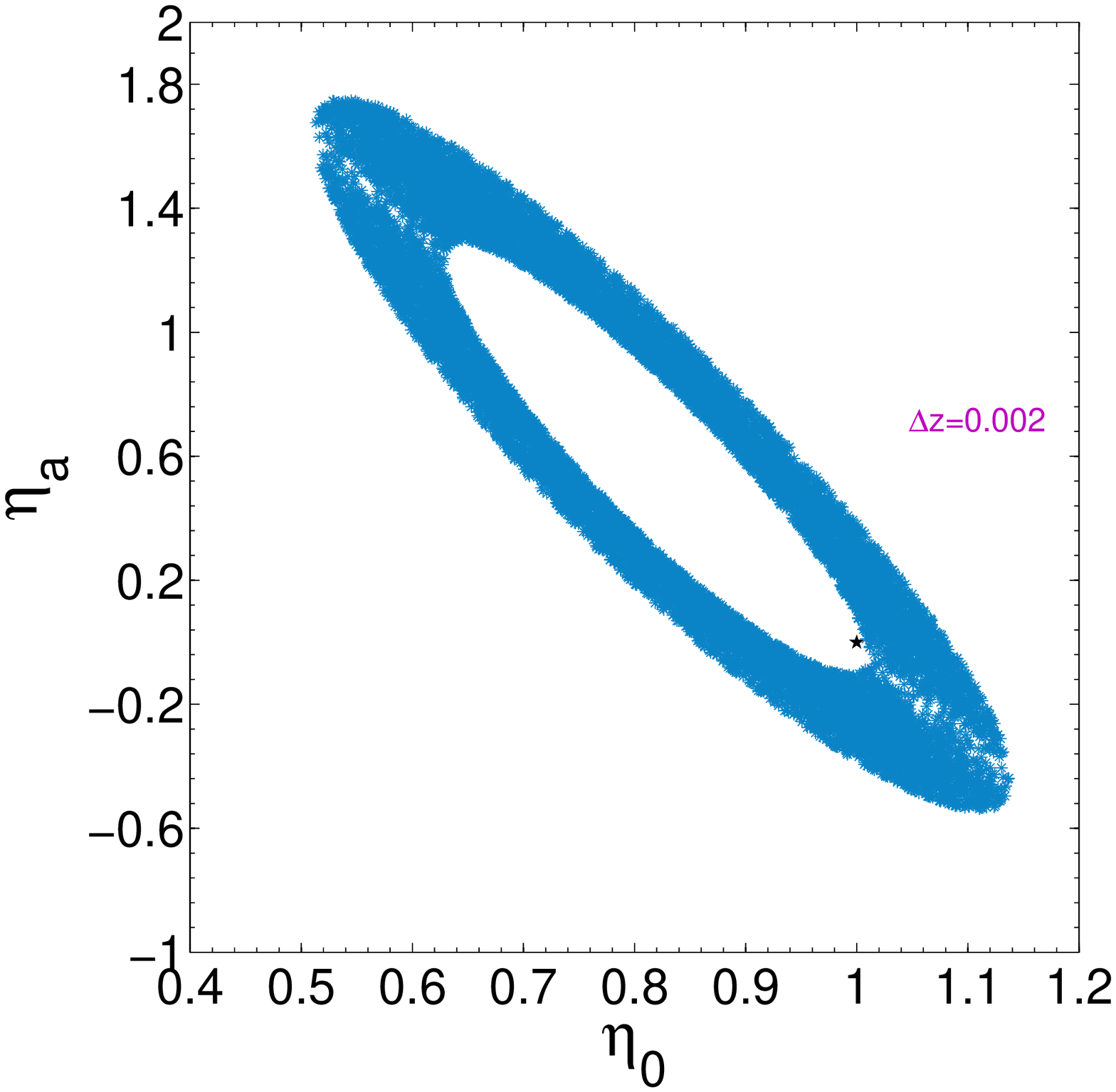}
 \end{center}

\caption{ Likelihood contours with the 33 ADDs of galaxy clusters
and the luminosity distances of the Union2 set at $1$ and $2\sigma$
CL for two-parameter forms in the $\eta_0-\eta_a$ plane [Left: for
$\eta(z)=\eta_0+\eta_1 z$; Middle: for
$\eta(z)=\eta_0+\eta_a\frac{z}{1+z}$; Right: for $\eta(z)=\eta_0 +
\eta_{a}\ln(1+z)$]. The filled stars represent the cases with no
violation of the DD relation ($\eta_0=1$ and $\eta_a=0$).
 \label{fig2}}
\end{figure}

The above analyses are based on the assumption that the
redshift-independent model parameter is a constant $\eta_{0}=1$. Now
we take it as a varying parameter to examine the DD relation by
assuming more general expressions: V. $\eta (z) = \eta_0 +
\eta_{a}z$; VI. $\eta (z) = \eta_0 + \eta_{a}z/(1+z)$; VII. $\eta
(z) = \eta_0 + \eta_{a}\ln (1+z)$. Fitting results from
two-parameter forms with the ADDs of galaxy clusters and the
luminosity distances of the Union2 set with $\Delta z=\sigma_{z,\rm
tot}=0.002$ are shown in Fig.~\ref{fig2} and summarized in Table
\ref{tab1}. Our results suggest that there is no violation of the DD
relation for two-parameter parameterizations at $1\sigma$ CL for
model V and VII, and at $2\sigma$ CL for model VI, which are more
stringent than those obtained in \citet{Li11}, where the DD relation
are consistent at $2\sigma$ CL for the spherical sample of galaxy
clusters.

From Fig.~\ref{fig1}-~\ref{fig2} and Table~\ref{tab1}, we can find
that the DD relation can be accommodated at $1\sigma$ CL for the
Bonamente et al. sample except model VI. Our results differs from
those obtained by \citet{HLR10}, where the results from the
Bonamente et al. sample give a clear violation of the DD relation.
However, these results are more stringent than those obtained by
\citet{Li11}, where the DD relation is accommodated at $2-3\sigma$
CL for the spherical sample of galaxy clusters.

\begin{figure}
\begin{center}
\includegraphics[width=0.33\hsize]{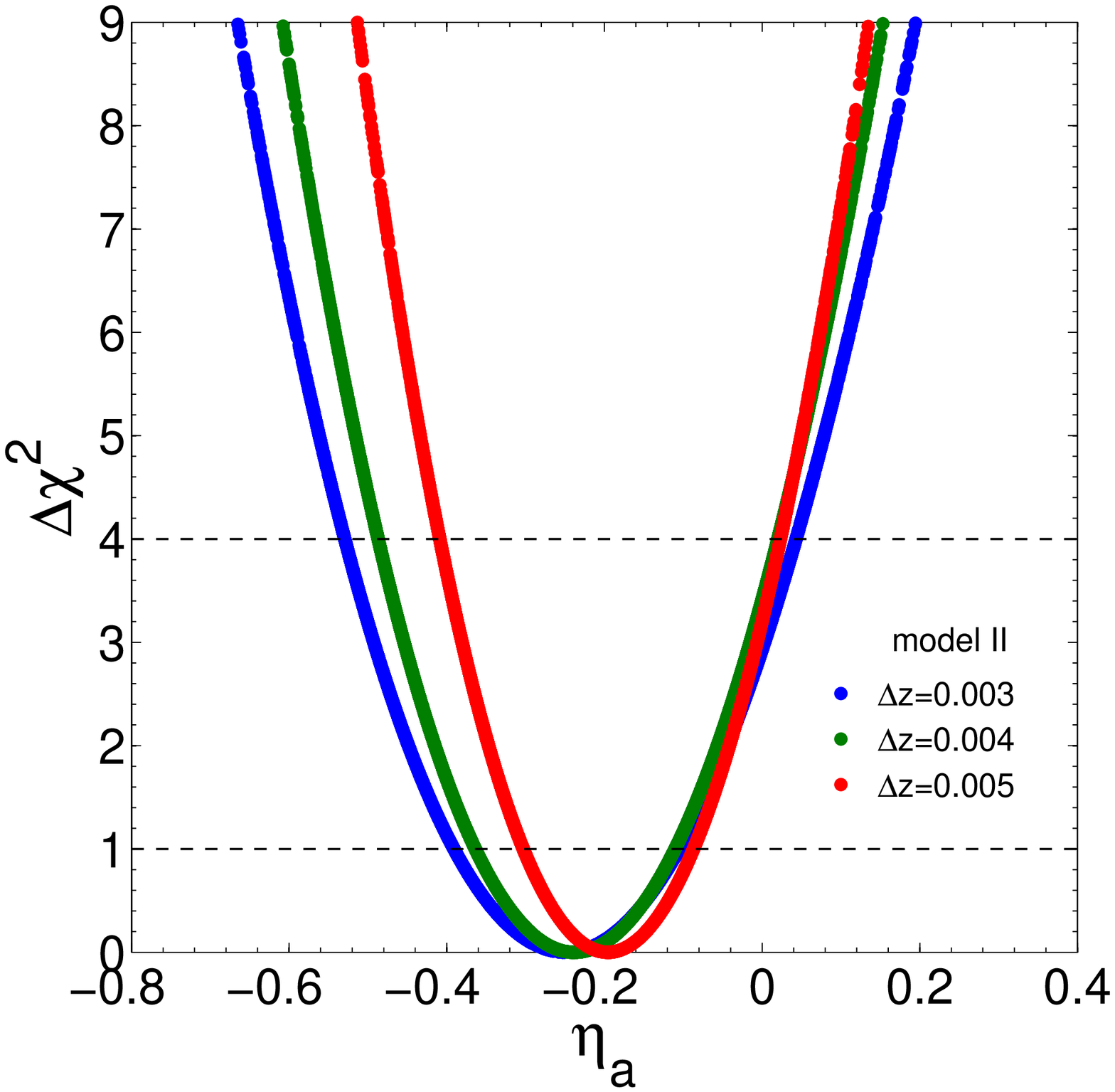}
\includegraphics[width=0.33\hsize]{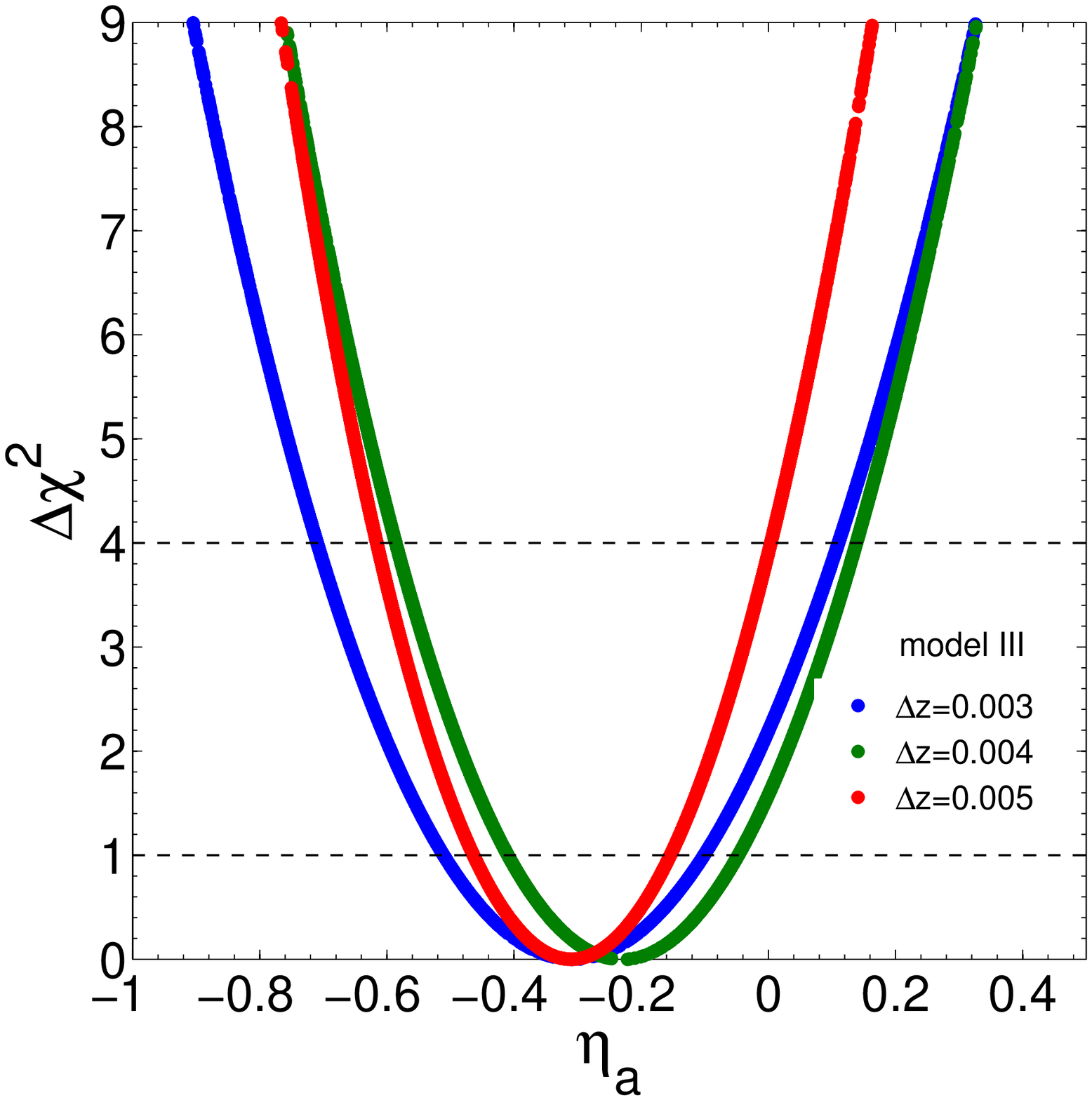}
\includegraphics[width=0.33\hsize]{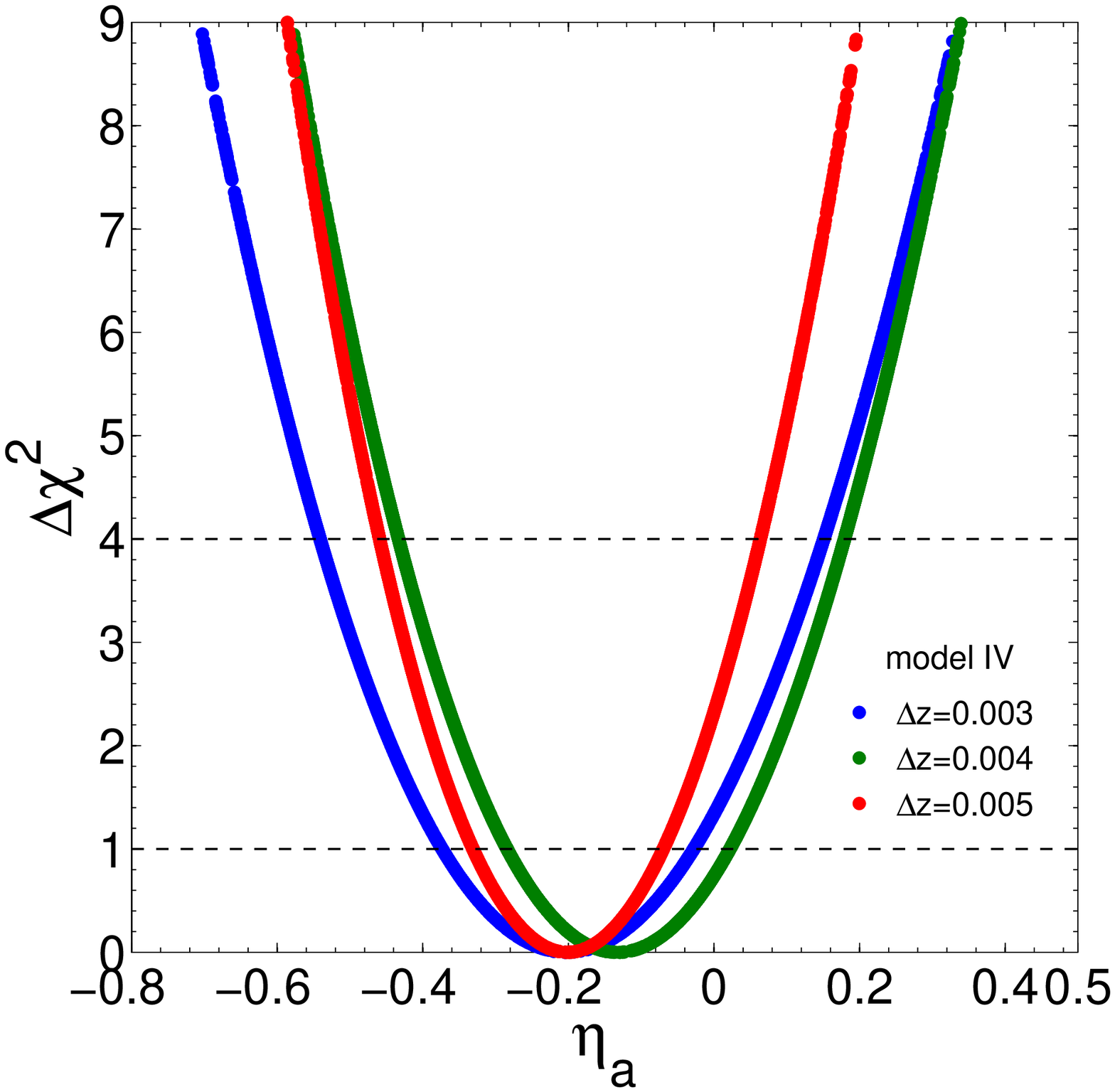}
\end {center}
 \caption{The likelihood distribution function in the $\eta_a-\Delta\chi^2$ planes for the three one-parameter forms of the
redshift (model II, III, IV) with varying $\Delta
z=0.003,0.004,0.005$. \label{fig3}}
\end{figure}

\begin{figure}
 \begin{center}
 \includegraphics[width=0.7\hsize]{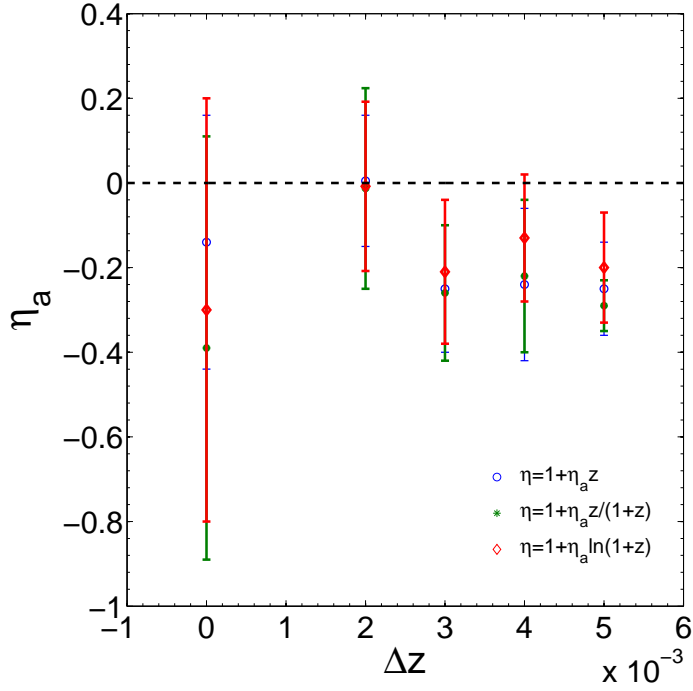}
 \end {center}
 \caption{The $1\sigma$ error bar of $\eta_a$ as a function of $\Delta z=0-0.005$ for the three one-parameter forms of the
redshift (model II, III, IV).
 \label{fig4}}
 \end{figure}

After identifying the constraints on $\eta$ obtained with the
minimum acceptable $\Delta z=0.002$, we may consider different
values of $\Delta z$ for examining the role of the choosing of
$\Delta z$  played in constraints. For the selection criteria of
$\Delta z=0.003,0.004,0.005$, there are 35, 37, 38 data pairs,
respectively. In Fig.~\ref{fig3}, we show the corresponding
constraints on $\eta_a$ for the three one-parameter forms of the
redshift: II. $\eta(z)=1+\eta_a z$; III.
$\eta(z)=1+\eta_a{z}/({1+z})$; and IV. $\eta
(z)=1+\eta_{a}\ln(1+z)$. Finally, we plot the $1\sigma$ error bar of
$\eta_a$ as a function of $\Delta z=0.002-0.005$ in Fig.~\ref{fig4}.
For comparison, we also show the results of $\Delta z=0$ with 14
data pairs.

From Fig.~\ref{fig3}-~\ref{fig4}, we can find that the choosing of
$\Delta z$ may play an important role in this model-independent
test, and the results for $\Delta z =0.005$ in our test show the DD
relation is ruled out at $2\sigma$ CL, which are close to those of
\citet{HLR10} where the DD relation is ruled out at $3\sigma$ CL,
and consistent with those obtained in \citet{Li11}, where the DD
relation are consistent at $3\sigma$ CL for the spherical sample of
galaxy clusters.

\section{Conclusions}\label{sec:Conclusions}

In this paper, we have discussed a new model-independent
cosmological test for the distance-duality relation, $\eta(z) =
D_{L}(1+z)^{-2}/D_{A}$. We consider the angular diameter distances
from galaxy clusters obtained by using Sunyaev-Zeldovich effect and
X-ray surface brightness together with the luminosity distances
given a sub-sample of SNe Ia taken from the Union2 data. The key
aspect is that SNe Ia is carefully chosen to have the minimum
acceptable redshift difference of the galaxy cluster ($\Delta
z=\sigma_{z,\rm tot}=\sigma_{z,\rm SN}+\sigma_{z,\rm cluster}$). For
the sake of generality, the $\eta$ parameter is also parameterized
in seven different forms, namely, four one-parameter models.: (I)
$\eta =\eta_{0}$, (II) $\eta = 1+\eta_{a}z$, (III)
$\eta=1+\eta_{a}z/(1+z)$, (IV) $\eta = 1+\eta_{a}\ln(1+z)$ and three
two-parameter models: (V) $\eta = \eta_{0}+\eta_{a}z$,
(VI)$\eta=\eta_{0}+\eta_{a}z/(1+z)$, (VII) $\eta =
\eta_{0}+\eta_{a}\ln(1+z)$.

By assuming $\eta$ a constant, we obtain $\eta_{0}=0.97^{+
0.05}_{-0.06}$ at 1$\sigma$. For the redshift-dependent
one-parameter forms of model II, model III, and model IV, we obtain
$\eta_a=-0.01^{+ 0.15}_{-0.16}$, $\eta_a=-0.01^{+ 0.21}_{- 0.24}$,
and $\eta_a=-0.01^{+ 0.22}_{- 0.19}$, respectively, which are well
consistent with no violation of the distance-duality relation. We
furthermore put forwards three kinds of two-parameter
parametrizations corresponding to model II, III and IV,
respectively. The standard values without any violation of the
reciprocity relation ($\eta_{0}=1$ and $\eta_{a}=0$) is still
included at $68.3\%$ CL (1$\sigma$) for model V, VII and at $95.8\%$
CL (2$\sigma$) for model VI. It is shown that there is no
conceivable evidence for variations of the duality distance relation
for the Bonamente et al. sample, since it is marginally satisfied
within $1\sigma$ CL for most cases, which is different from those
obtained by \citet{HLR10}, where the results from the Bonamente et
al. sample give a clear violation of the DD relation, and more
stringent than those obtained by \citet{Li11}. By further
considering different values of the redshift difference $\Delta z$,
we find that the choosing $\Delta z$ may play an important role in
this model-independent cosmological test of the DD relation, and the
results for $\Delta z =0.005$ in our test  show the DD relation is
ruled out at $2\sigma$ CL, which are close to those of \citet{HLR10}
where the DD relation is ruled out at $3\sigma$ CL, and consistent
with those obtained in \citet{Li11}, where the DD relation are
consistent at $3\sigma$ CL for the spherical sample of galaxy
clusters.

It is still interesting to see whether those conclusions may be
changed with larger SNe Ia and clusters data in the future, which
reinforces the interest in the observational search for more samples
of galaxy clusters with smaller statistical and systematic
uncertainties, as well as the determination of their angular
diameters trough the combination of SZE and X-ray surface
brightness.

\begin{acknowledgements}
We thank Prof. Zong-Hong Zhu, Dr. Yi Zhang, as well as Heng Yu,
XiaoLong Gong, Hao Wang, Xing-Jing Zhu, Yan Dai, Yu Pan, Fang Huang,
Jing Ming, Kai Liao for discussions. This work was supported by the
National Natural Science Foundation of China under the Distinguished
Young Scholar Grant 10825313 and Grant 11073005, the Ministry of
Science and Technology national basic science Program (Project 973)
under Grant No.2007CB815401, the Fundamental Research Funds for the
Central Universities and Scientific Research Foundation of Beijing
Normal University.
\end{acknowledgements}

\end{document}